**Broken Symmetry Quantum Hall States in Dual Gated ABA Trilayer Grapehene**

Yongjin Lee[1], Jairo Velasco Jr.[1], David Tran[1], Fan Zhang[2], W. Bao[1], Lei Jing[1], Kevin Myhro[1], Dmitry Smirnov[3], Chun Ning Lau[1*]

[1]Department of Physics and Astronomy, University of California, Riverside, Riverside, CA 92521

[2]Department of Physics and Astronomy, University of Pennsylvania, Philadelphia, PA 19104

[3]National High Magnetic Field Laboratory, Tallahassee, FL 32310

**Abstract**

We present low temperature transport measurements on dual-gated suspended trilayer graphene in the quantum Hall (QH) regime. We observe QH plateaus at filling factors $v$=-8, -2, 2, 6, and 10, in agreement with the full-parameter tight binding calculations. In high magnetic fields, odd-integer plateaus are also resolved, indicating almost complete lifting of the 12-fold degeneracy of the lowest Landau levels (LL). Under an out-of-plane electric field $E_\perp$, we observe degeneracy breaking and transitions between QH plateaus. Interestingly, depending on its direction, $E_\perp$ *selectively* breaks the LL degeneracies in the electron-doped or hole-doped regimes. Our results underscore the rich interaction-induced phenomena in trilayer graphene.

[*] Email: lau@physics.ucr.edu

As a fascinating two-dimensional (2D) system with chiral charge carriers and spectacular electronic, mechanical and thermal properties, graphene and its multilayer counterparts[1-3] have emerged as new platforms for investigation of QH physics. A number of novel phenomena have been observed, such as multicomponent fractional QH effect in monolayer graphene(MLG)[4-7], insulating $v$=0 states in MLG and bilayer graphene(BLG)[8-13], electric field-driven transitions among symmetry-broken QH states in BLG[14-16], and chiral charge carriers with berry phase of $3\pi$[17] and Lifshitz transition in ABC-stacked trilayer graphene (TLG)[18].

Like MLG and BLG, TLG offers an exciting platform with unique band structures. In particular, ABA-stacked TLG hosts mirror symmetry with respect to the middle layer (Fig. 1a inset), and its band structure can be viewed as a combination of the linear dispersion of MLG and parabolic dispersion of BLG, *i.e.* the so-called "2+1" model within tight-binding calculations[19-27] (Fig. 1a-b). Though TLG has attracted more attention recently[17, 18, 28-34], the nature of broken symmetry QH states in TLG and their evolutions under electric and magnetic fields remain experimentally unexplored.

In this Letter we report transport measurements on high mobility dual-gated ABA TLG devices in the QH regime. At low magnetic field $B$<4T, we resolve single particle QH states at filling factors $v$=-8, -2, 2, 6, and 10, which can be accounted for by the "2+1" tight-binding model that includes all hopping parameters[35]. At higher $B$, we observe additional states at $v=\pm1$, $\pm3$, -4 and -5, indicating almost complete lifting of the degeneracy of the lowest LL. At constant $B$, application of an out-of-plane electric field $E_\perp$ gives rise to degeneracy breaking and transitions between QH plateaus, suggesting the interplay of layer polarization induced by $E_\perp$ and $B$-enhanced exchange interactions of these states. Finally, depending on its polarity, we find the $E_\perp$ *selectively* breaks the LL degeneracy in the electron-doped or hole-doped regimes.

TLG sheets are isolated via mechanical exfoliation on Si/SiO$_2$ substrates, identified by optical contrast and Raman spectroscopy[36], and coupled to Cr/Au electrodes and Cr suspended top gates[37, 38]. The devices are completed by removal of SiO$_2$ under the graphene with HF etching (Fig. 1b). All data are taken at 300mK in He$^3$ refrigerators, and similar phenomena are observed in 3 devices.

An important advantage of the suspended top-gate structure is its compatibility with post-fabrication annealing that may dramatically improve sample quality. Fig. 1c displays the two-terminal conductance $G$ as a function of back gate $V_{bg}$ before (blue curve) and after (red curve) current annealing. After annealing, the curve becomes 'V'-shaped, with charge neutrality close to zero, drastically lower minimum conductance, and high field effect mobility of ~15,000 cm$^2$/Vs.

In the simplest tight binding model that includes only the nearest neighbor in-plane and inter-plane hopping parameters $\gamma_0$ and $\gamma_1$, the band structure of ABA-stacked TLG consists of the MLG-like and BLG-like branches touching at a single point (Fig. 1a, left panel). In sufficiently large applied $B$, the charges' cyclotron orbits coalesce to form discrete LLs, with energy given by[19, 39-41]:

$$E_{M,N} = \pm\sqrt{2\hbar v_F^2 eB|N|} \quad \text{and} \quad E_{B,N} = \pm\frac{\hbar eB}{m^*}\sqrt{N(N-1)} \tag{1}$$

The lowest LL is 12-fold degenerate, giving rise to quantized plateaus at filling factors $\nu = nh/Be = \ldots -10, -6, 6, 10, 14\ldots$. Here $e$ is electron charge, $n$ the induced charge density, $h$ Planck's constant, $v_F \sim 10^6$ m/s the Fermi velocity, $m^* = \frac{\gamma_1}{\sqrt{2}v_F^2} \sim 0.02\text{-}0.04\, m_e$, $m_e$ the electron rest mass, $\gamma_1 \sim 0.3$ eV is the interlayer coupling, and $N$ is an integer denoting the LL index. Fig. 2a shows the standard LL "fan diagram" of the device, i.e. $G$ (color scale) as a function of $V_{bg}$ (horizontal axis) and $B$ (vertical axis). The QH plateaus appear as the colored bands that diverge from $B=0$ and the charge neutrality point (CNP). From the fan diagram, the back gate's coupling efficiency is estimated to be $\alpha_{bg} \sim 3.8 \times 10^{10}$ cm$^{-2}$/V.

To accentuate the evolution of the QH plateaus with $V_{bg}$ and $B$, we plot $dG/dV_{bg}(V_{bg}, B)$ of the same data set in Fig. 2b. The filling factor of each plateau, which appears as a white band, $\nu = nh/Be = \alpha_{bg}V_{bg}h/Be$, is calculated from its slope in the $V_{bg}$-$B$ plane and labeled in Fig. 2b. The most prominent feature is the very strong $\nu=-2$ plateau in the hole-doped regime, which is resolved at $B$ as small as 0.25T. (Here we define hole-doped and electron-doped regime to have negative and positive filling factors, respectively.) Line traces $G(V_{bg})$ at several $B$ values for $B<4.2$T are shown in Fig. 2c. When replotted as a function of $\nu$, the traces nearly collapse into a single curve, with properly quantized plateaus at $\nu=-2, 2, 6$ and $10$.

The appearance of robust $\nu=6$ and 10 states agrees with Eq. (1) as well as prior reports[30, 32, 34]. On the other hand, our observation of the $\nu=2$ and in particular the exceedingly robust $\nu=-2$ plateaus, is unexpected from Eq. (1). This can however be accounted for by the "2+1" model that takes remote hopping into account – instead of MLG-like and BLG-like bands both touching at a single point, including next-nearest hopping parameters ($\gamma_2$ and $\gamma_5$) leads to bands that are individually gapped, with a relative vertical offset between the MLG-like and BLG-like bands, whose tops of valence bands are located at $-\gamma_2/2$ and $\gamma_2/2$, respectively (Fig. 1a, right panel). Consequently, the LL spectrum of such a band structure is modified from Eq. (1) as follows: (i). since ABA stacked TLG obeys mirror symmetry but not inversion symmetry[35], its valley degeneracy is not protected; the broken valley degeneracy of the lowest LL[35] manifests as $\nu=\pm2$ plateaus, as observed experimentally; (ii). the spectrum is particle-hole asymmetric, and (iii). LLs originating from the MLG-like and BLG-like bands cross at energy $\sim\pm\gamma_2/2$.

All three features are observed in our experimental data. Apart from the robust $\nu=\pm2$ plateaus, the particle-hole asymmetry is clearly reflected in the sequence of resolved plateaus -- the $\nu=6$ and 10 plateau is observed only in the electron doped regime, and $\nu=-8$ solely in the hole-doped regime. The dark blue feature at $V_{bg}\sim -5$V, indicated by the dotted circle in Fig. 2b, corresponds to the crossings between LLs that belong to the MLG and BLG-like spectra. From the data, the crossings occur at $\sim-1.9\times10^{11}$ cm$^{-2}$, corresponding to $\sim-8$ meV. Thus our data suggest $\gamma_2\sim-16$ meV in TLG, in reasonable agreement with the value from bulk graphite, $-20$ meV[42].

Thus far the $v=-2, 2, 6$, and 10 plateaus are well accounted for by single particle tight binding calculations. At larger $B$, we also observe additional plateaus at $v=\pm 1, \pm 3, -4$ and $-5$, which indicate almost complete lifting of spin, valley and orbital degeneracies in the lowest LL. The $v=0$ plateau, although resolved, is $\sim 0.3\ e^2/h$ at 18T. This lack of true insulating behavior is likely due to the presence of small amount of residual impurities. Fig. 2c-d plots $G(V_{bg})$ and $G(v)$ at $B=4.5, 6, 7, 8$ and 10T, respectively, showing excellent conductance quantization. The $v=\pm 1$ plateaus are resolved at $B$ as low as 4.5T, and persists to 18T (Fig. 2e-f), the highest available field. These additional plateaus cannot be accounted for by any tight binding model, and their appearance at high $B$ values in samples with high mobility ($\geq 10,000$ cm$^2$/Vs) strongly suggests symmetry breaking arising from electronic interactions. In fact, they can be qualitatively understood in terms of QH ferromagnetism and Hunds rule-like filling of the 12-fold degenerate lowest LL[43]. Within this model, the LLs between $v=-6$ and 6 are filled in the order of maximizing spin, chirality (BLG-like branch first), valley and orbital indices. At large $B$, the $v=-5, -4, -3, -2, 1, 2$, and 3 states belong to the BLG-like branch, while the $v=-1, 0, +4$ and $+5$ states to the MLG-like branch[43]. As observed experimentally, all the BLG-like states are fully resolved, whereas only the $v=-1$ (and to some extent the $v=0$) state in the SLG-like branches is observed. This is consistent with previous observations that the QH ferromagnetic states in BLG are more easily resolved, due to its enhanced density of states and stronger electronic interactions near charge neutrality.

We now focus on the QH states in the presence of both top and back gates. Sweeping both top and back gate voltages enables independent modulation of the electric field $E_\perp$ and total charge carrier density $n$ in TLG, which has emerged as a critical tool to study the broken symmetry states in bilayer graphene[14, 15]. For ABA-stacked TLG, $E_\perp$ breaks its mirror reflection symmetry, and is expected to give rise to otherwise unresolved plateaus or the stabilization of existing plateau with finite $E_\perp$.

In Fig. 3a-c $G$ (color scale) is plotted as a function of $E_\perp$ (vertical axis) and $n$ (horizontal axis) at $B=5.5, 8$ and 14T, respectively. The vertical color bands correspond to the conductance plateaus at different filling factors. Fig. 3d plots $G(n)$ at $B=8$T and $E_\perp=0, 43$ and 73 mV/nm, respectively. At $E_\perp=0$, plateaus $v=0, 1, 2$ and 3 are observed. At $E_\perp=43$ mV/nm, the first 3 plateaus remain relatively unchanged, whereas the $v=3$ plateau is better resolved, and the $v=4$ plateau emerges. Thus, our data suggest that layer polarization is an important component in the $v=3$ and 4 states.

Another striking feature in these $G(n, E_\perp)$ plots is the dependence of the $v=0$ plateau on $E_\perp$: it abruptly increases from 0.3 to $\sim 1\ e^2/h$ at a critical $E_{\perp c}$ value, and decreases to 0.3 again for a larger $E_\perp$ (Fig. 3d). The $G(n)$ trace at $E_{\perp c}$ is characterized by the absence of the $v=0$ plateau (Fig. 3e, blue line). Taken together, our data suggest a transition from spin-polarized to layer-polarized states driven by $B$ and $E_\perp$, respectively, and is highly reminiscent of that in BLG[14, 15].

To further investigate the transition between the LLs, we examine the dependence of $E_{\perp c}$ on $B$. In BLG, $E_{\perp c}$ is linearly dependent on $B$, with a slope ~ 13 mV/nm/T and extrapolates to a finite value ~ 12 mV/nm at $B=0$[14, 15]. Fig. 3f plots $E_{\perp c}$ vs. $B$ for 3 different devices (device D1 was measured at 2 separate locations). For $B<8T$, within the scatter in the data, $E_{\perp c}$ is approximately linear in $B$, with a best-fit equation $E_{\perp c}$ (mV/nm) = 19.7 + 6.9$B$. The finite intercept indicates the evolution of subband gaps at B=0 when mirror symmetry is broken by electric fields, or alternatively, a transition from spin-polarized to layer-polarized states that persists to zero magnetic field. Interestingly, when $B$ is extended to 18T, the data points are no longer linear; instead, they can be adequately fitted to the equation $E_{\perp c}$ (mV/nm) = 8.3 + 21.7 $B^{1/2}$, suggesting that Coulomb interactions play an essential role at large $B$. There is little theoretical work on LL transitions in ABA-stacked TLG in the presence of electric and magnetic fields, and this phenomenon warrants further experimental and theoretical investigation.

Finally, we focus on a peculiar feature of the conductance in the presence of $E_\perp$ and $B$. Fig. 4 plots $G(E_\perp,n)$ at $B=7T$. At finite $E$, the $G(n)$ traces are asymmetric with respect to electrons and holes. Interestingly, such asymmetry depends on the direction of the applied $E_\perp$, and reverses upon reversal of the sign of $E_\perp$. In Fig. 4a, this asymmetry can be seen as the asymmetric appearance of the bright blue band to the right (left) of the charge neutrality point for positive (negative) $E_\perp$. Fig. 4b plots the $G(n)$ curves at $E_\perp=0$, -17 and 13.6 mV/nm, respectively. The $\nu=$-1 plateau was only resolved for $E_\perp<0$, whereas the $\nu=1$ state was better resolved for $E_\perp>0$. Thus, $E_\perp$ appears to selectively break the symmetry of LLs of the electron- or hole- doped regimes, depending on its polarity. We currently do not have an explanation for this phenomenon. It may be related to the particle-hole asymmetry in few layer graphene's band structure[44], or to more intriguing phenomena such as spin-orbit interactions or magento-electric effects[45, 46]. Further experimental investigation will be necessary to fully elucidate its origin.

In conclusion, using dual-gated high mobility samples, we observe several intriguing phenomena related to the broken symmetry QH states in ABA-stacked TLG, including almost complete lifting of the spin, valley and orbital degenerancies of the lowest LL, stabilization of some of these states by $E_\perp$, transition between LLs driven by $E_\perp$ and $B$, and a particle-hole asymmetry that depends on the polarity of $E_\perp$. Our study demonstrates the rich interaction physics in ABA TLG in the $E_\perp$-$B$-$n$ phase space. A number of unresolved questions, such as the dependence of $E_{\perp c}$ on $B$ at large field, and dependence of the electron-hole asymmetry on $E_\perp$, await further studies.

We thank Yafis Barlas for helpful discussions. This work was supported in part by NSF CAREER DMR/0748910, NSF/1106358 and the FENA Focus Center. F.Z. was supported by DARPA SPAWAR N66001-11-1-4110. Part of this work was performed at NHMFL that is supported by NSF/DMR-0654118, the State of Florida, and DOE. This material is partly based on research sponsored by DARPA/DMEA H94003-10-2-1003. The US Government is authorized to reproduce and distribute reprints for Government purposes, notwithstanding any copyright notation thereon.

**Figure 1.** (a). Low energy band structure of ABA-stacked TLG calculating (left panel) using only $\gamma_0$ and $\gamma_1$, (right panel) using $\gamma_0 - \gamma_5$. Inset: ABA-stacked TLG lattice with hopping parameters $\gamma_1$-$\gamma_5$. (b). SEM image of a dual-gated suspended TLG device. (c) $G(V_{bg})$ before(blue) and after(red) current annealing.

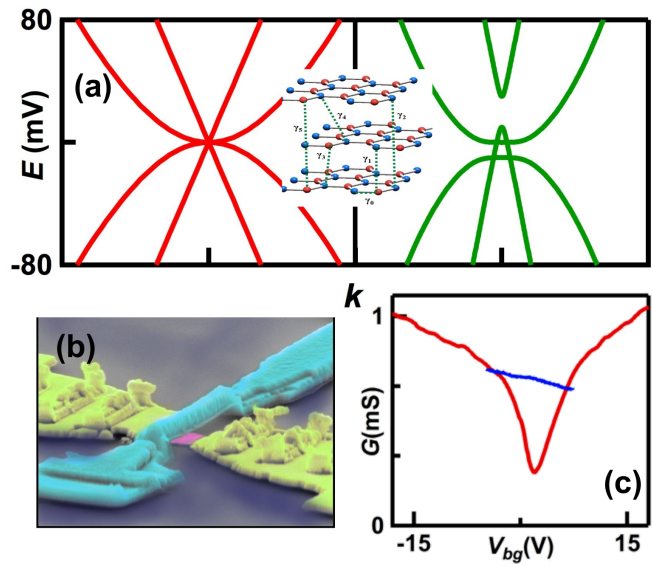

Figure 2. (a-b). $G(V_{bg}, B)$ and $dG/dV_{bg}$ of a TLG device. Numbers indicate filling factors. (c). $G(V_{bg})$ and $G(\nu)$ at $B$=1.5, 2.2, 3.5 and 4.2 T, respectively. (d). $G(V_{bg})$ and $G(\nu)$ at $B$=4.5, 6, 7, 8 and 10 T. (e). $G(V_{bg}, B)$ and $G(\nu)$ at $B$=10, 12, 14, 16 and 18T.

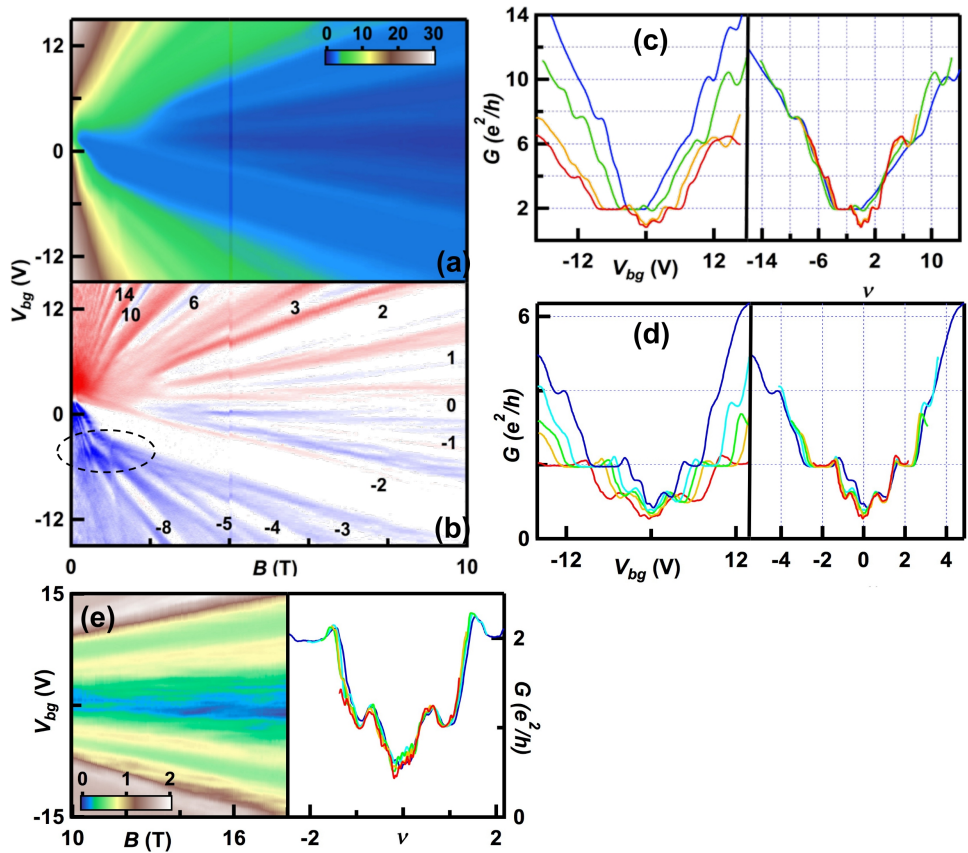

Figure 3. (a-c). $G(n, E_\perp)$ at $B$=5.5, 14 and 8T. (d). $G(n)$ along the horizontal lines in (c) at $E_\perp$=0 (red), 43 (green dotted line) and 73 mV/nm (blue), respectively. (e). $G(E_\perp)$ along the vertical line in (c) at $n$=0. (f). $E_{\perp c}(B)$ from 3 different devices. The black and orange lines correspond to linear and $B^{1/2}$ fits, respectively.

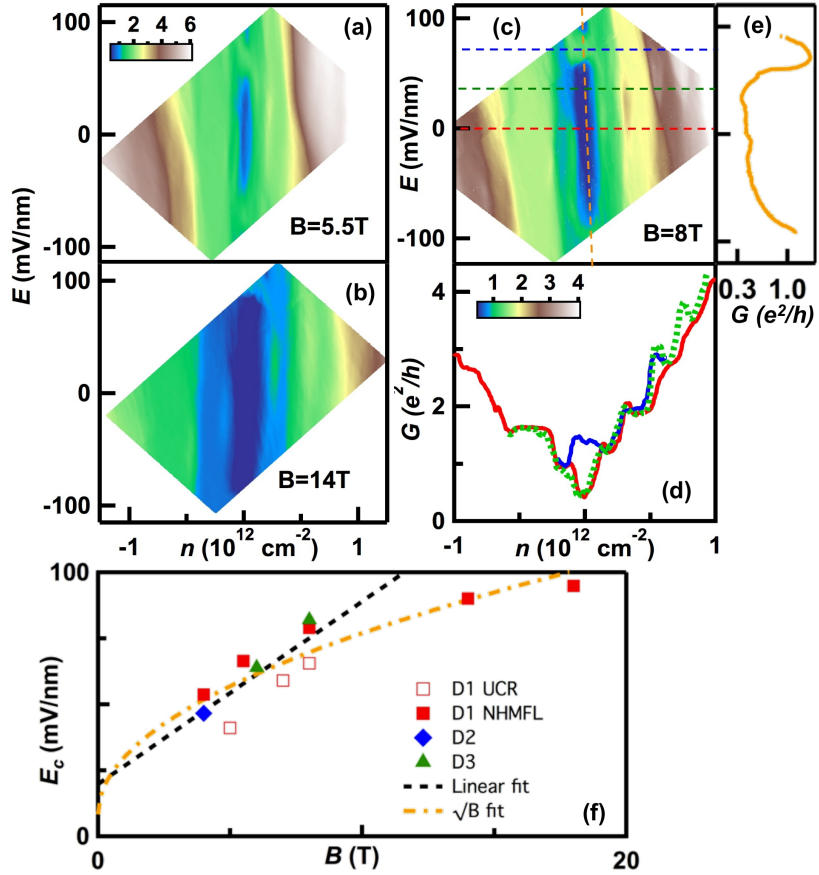

Figure 4. (a). $G(n, E_\perp)$ at $B=7$T. (b). $G(n)$ at $E_\perp=0$ (red solid line), -17 (green dotted line) and 13.6 mV/nm (blue dashed line).

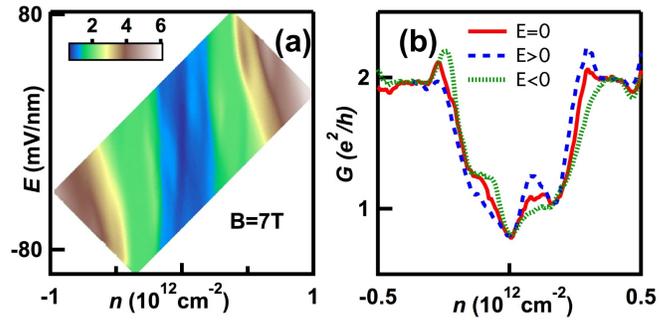